\documentclass[conference]{IEEEtran}

\IEEEoverridecommandlockouts
\usepackage{cite}
\usepackage{amsmath,amssymb,amsfonts}
\usepackage{algorithmic}
\usepackage{graphicx}
\usepackage{textcomp}
\usepackage{tabularx, booktabs, fancyvrb, pifont}
\usepackage{xcolor}
\def\BibTeX{{\rm B\kern-.05em{\sc i\kern-.025em b}\kern-.08em
    T\kern-.1667em\lower.7ex\hbox{E}\kern-.125emX}}
\begin{document}

\title{Towards Noise Robust Trigger-Word Detection with Contrastive Learning Pre-Task for Fast On-Boarding of New Trigger-Words\\
}

\author{\IEEEauthorblockN{Sivakumar Balasubramanian}
\IEEEauthorblockA{\textit{Samsung Research America} \\
siv.bala@samsung.com}
\and
\IEEEauthorblockN{Aditya Jajodia}
\IEEEauthorblockA{\textit{Samsung Research America} \\
aditya.2@samsung.com}
\and
\IEEEauthorblockN{Gowtham Srinivasan}
\IEEEauthorblockA{\textit{Samsung Research America} \\
g.srinivasan@samsung.com}
}

\maketitle

\begin{abstract}
Trigger-word detection plays an important role as the entry point of user's communication with voice assistants. But supporting a particular word as a trigger-word involves huge amount of data collection, augmentation and labelling for that word. This makes supporting new trigger-words a tedious and time consuming process. To combat this, we explore the use of contrastive learning as a pre-training task that helps the detection model to generalize to different words and noise conditions. We explore supervised contrastive techniques and also propose a novel self-supervised training technique using chunked words from long sentence audios. We show that both supervised and the new self-supervised contrastive pre-training techniques have comparable results to a traditional classification pre-training on new trigger words with less data availability. 

\end{abstract}


\section{Introduction}
Trigger-word detection holds an important role in various speech based applications. Voice assistants such as Amazon's Alexa and Samsung's Bixby use it as a wake-word detection module for enabling users to start interacting with the assistant. Training wake-word detection model such as \cite{C1,C2,C3, C4} requires huge amount of utterances of the wake-word. The data also needs to encompass variations in accents, genders and background noises. Thus training a wake-word detection model for supporting a particular wake-word requires great amount of utterance collection and augmentation \cite{C5,C6} for that particular word. It also requires proper labelling of the data to avoid false positives/negatives that could degrade training process \cite{C7}. Since the wake-word detection models are deployed on-device to protect user privacy, it imposes constraints on model size to aid in low latency and power consumption.  The above mentioned factors, make it difficult for supporting new wake-words in voice assistants.  

Enabling faster support of new wake-words require a training technique that can learn the context of the detection task from generic open datasets and adapt to new trigger-words with less labelled data and data augmentation. Transfer learning techniques such as \cite{C6,C8} use an ASR task for normalizing the weights of the trigger-word detection model. Also a multi-class classification model as shown in \cite{C9,C10} can be trained and the weights can be used to initialize a model for a particular classification task with less labelled data. However, these technique requires huge amount of labelled data for multiple trigger-words or large sets of ASR Corpus with transcriptions.

Contrastive Learning is a training technique that tries to learn an embedding space where similar data samples are closer and dissimilar samples are farther away from each other. It involves training using pairs of similar and dissimilar datasets using a contrastive loss function as in \cite{C11}. In recent works, this contrastive loss has been used widely as a pre-task to a main task \cite{C12,C14,C15,C16,C17,C18,C19,C20,C21}. The pre-task can be a supervised \cite{C12, C17,C21} or a self-supervised \cite{C14,C15,C16,C18,C19,C20} task. 

Recent results have shown that a supervised contrastive loss can be used for getting better performance in various classification tasks in speech based applications \cite{C17,C21}. But this requires label information of the data.  Self-supervised contrastive learning tries to learn representations without labels but has a degraded performance due to hard negative training examples \cite{C13}.

In this work, we explore the transfer learning ability of contrastive learning pre-task for wake-word detection and compare the performance with traditional transfer learning techniques. The wake-word detection task (main task) is performed on words that were not used for training the pre-task and without noise augmentations. This is done to see how well the contrastive pre-task generalizes to background noise. Specifically, we try the following contrastive pre-task models.

\begin{itemize}

\item A supervised contrastive model trained using pairs of single word utterances.
\item  A pseudo supervised contrastive model wherein trigger-word utterances are divided into two sets using labels and contrastive task is trained with pairs from the two sets without labels.
\item A self-supervised contrastive model trained with a novel training scheme using chunked utterances from long sentence audios.
\end{itemize}

\section{Experimental Setup}

To explore the transfer learning and adaptation capability of contrastive learning pre-task we train various models with different pairing strategies. We also train a traditional transfer learning model with a multi-class classification pre-task. The main task is either a one vs all classification task for various trigger-words or a multi-class trigger-word classification task. 

\subsection{Datasets and Augmentation}

We use the Google Speech Command (GSC) V2 dataset \cite{C22}  as the single word utterance data. The GSC dataset contains 35 trigger-word classes. Each trigger-word class has utterances of that word by multiple speakers. We also use LibriSpeech \cite{C23}  as the long sentence audio dataset. Noise augmentations such as babble, music, car and cafe are performed on the GSC and chunked LibriSpeech data. We augment the noises at SNR levels ranging from 10 to 25. Apart from the noise augmentations we also include pitch shift and time shift augmentations in the positive audios pairs of the contrastive models.   We use 40 dimensional MFCC features at a sampling rate of 16000 Hz with a hamming window of width 25 ms and step 10 ms as input for the contrastive learning task as well as the classification task.  
\begin{table}[htbp]
\caption{Various datasets used for the pre-tasks.}
\label{tab:fdroid}
\normalsize
\centering
\begin{tabular}{| >{\centering} m{2.2 cm}| m{5.2cm}|} 
\hline
\textbf{Exp} &
\textbf{Pre-task} \\
\hline
Classification random init (C) & None  \\
\hline
Classification Transfer (CT1) &
Classification task on 30 trigger-words \\
\hline
Classification Transfer (CT2) &
Classification task on 15 trigger-words \\
\hline
Supervised Contrastive Transfer (SC1) &
Positive pairs of same word (clean + aug) and negative pairs of  different words (clean + aug) for 30 trigger-words. \\
\hline
Supervised Contrastive Transfer (SC2) &
Positive pairs of same word (clean + aug) and negative pairs of  different words (clean + aug) for 15 trigger-words.\\
\hline
Self-Supervised Contrastive Transfer (SSC) &
Positive pairs of utterance from set-1 and negative pair of set-1 and set-2  from chunked LibriSpeech. Set-1 and Set-2 contain randomly selected 30\% and 70 \% chunked audios respectively.  \\
\hline
Pseudo-Supervised Contrastive Transfer (PSC) &
Positive pairs of utterance from set-1  and negative pair of set-1 and set-2 utterances from GSC data. set-1 and set-2 contain utterances from 15 trigger-word each  \\
\hline
Self-Supervised Contrastive Transfer with Hard Negatives (SSHN) &
Positive pairs made of an utterance and its augmentation and negative pairs chosen randomly from all other utterances in batch of utterances. Utterances from 30 GSC trigger-words are used.  \\
\hline
\end{tabular}
\end{table}
\subsection{Contrastive Loss}
We use a siamese architecture as in \cite{C11} for the contrastive learning task. Pairs of utterances are given as inputs to the network and the final dense layer representation for each utterance is used for calculating the contrastive loss.

We use a negative exponential manhattan distance for the contrastive loss function as shown in Equation 1.

\begin{align} 
  D &= e^{-\sum_{i=0}^{n} \lvert left_i-right_i \rvert} 
 \end{align}
 where left and right are the final dense representation of utterance 1 and utterance 2 of the pair of utterances and n is the dense vector dimension. The exponential of the negative manhattan distance between the representations of the pairs of audios gives a number ranging from 0 to 1. A smaller value here corresponds to a larger distance. We use binary cross-entropy as the loss where the positive and negative pairs have labels of 0 and 1 respectively.

\subsection{Model Architecture}
We use a CNN ResNet Architecture \cite{C24}  with separable convolutions as the base model for all the experiments. We use six convolution layers with layer normalization followed by a flatten and two dense layers. The final dense layer is a vector with 128 dimension.  The experiments are performed by adding either one more dense layer in case of Classification models or a distance calculation layer in case of siamese task.

\subsection{Classification Models}

We train two types of classification models. One is a randomly initialized classification model (C) without a pre-task. This model is used as a baseline for comparing all other models. The other model is a traditional classification transfer (CT) model as in \cite{C9,C10}.  The GSC dataset is used for training the classification models. The single word utterances in GSC are divided into pre-task and main task trigger-word datasets. The main task dataset is not used for the pre-task training and thus considered as unseen wake-words during the main task.

We train the CT model for a multi-class classification pre-task with the pre-task dataset. CT1 and CT2 models represent a pre-task training with 30 and 15 trigger-words respectively. Then we replace the final dense layer of this model to correspond to a one vs all classification task for the main task. We additionally perform a multi-class trigger-word detection main task using the main task dataset. For the main task, we train dense layers of the models. We also try a fine-tuning experiment where we train the entire model for few epochs with a reduced learning rate.

\subsection{Supervised Contrastive Models}

Supervised contrastive training use the labels of the data for a particular task for determining positive (similar) pairs and negative (dissimilar) pairs. The positive pairs are formed by either pairing with other data having the same labels or by augmenting the same data sample. The negative pairs are formed by pairing clean and augmented data samples with different labels.

As mentioned above, we use pairs of utterances from same trigger-word class as well as augmentations of same utterance as positive pairs. Negative pairs consist pairs of utterances from different trigger-word class where the pairs are formed with both clean and augmented data.

In this work we use the GSC dataset with labels for the supervised contrastive (SC) model training. The GSC data is divided similar to the classification model training mentioned in previous subsection. The pre-task trigger-word dataset is used for training the contrastive task and main task trigger-word dataset is used for training the classification main tasks. The SC1 and SC2 represent a pre-task training with 30 and 15 trigger-words respectively. The final layer of the contrastive model is a distance calculation layer that implements the exponential negative manhattan distance between pairs of utterances. Similar to the classification transfer model, we replace the final layer and train the dense layers for the main tasks. We also try the fine-tuning experiment with reduced learning rate.

\subsection{Self-Supervised Contrastive Models}

Although supervised contrastive models are shown to have improved performance compared to other techniques, they still require labels of the training data. With self-supervised training technique we try to leverage context rich model weights from a given dataset without the labels. In general, self-supervised contrastive learning techniques try to pair each utterance with every other utterance in the batch. This makes sure that every class of data has a negative pair from every other class of data, but it also gets hard negatives from the same class of data. In this work, we propose a pseudo-supervised (PSC) and a new self-supervised (SSC) training technique to reduce the hard negative problem in the conventional self-supervised techniques. 

First, we train a conventional self-supervised model with hard negatives (SSHN). We make the positive and negative pairs from random batches of single word pre-task GSC utterances. The positive pairs are formed with an utterance and augmentations of the same utterance. The negative pairs for an utterance is chosen randomly from all the other utterances within the batch. This results in the possibility of hard negatives where different utterances with the same word can be labelled as a negative pair. 

Next, we train a model with a Pseudo-supervised approach (PSC).  In this approach we separate the pre-task GSC dataset into 2 sets using labels. We use the random utterances from set-1 and augmentations of the same utterance as positive pairs. We form negative pairs with clean utterances from set-1 and clean/augmented utterances from set-2. There is no pairing between different utterances within set-1 and set-2 utterances are only used for negative pairing with set-1 utterances. By doing this we make sure that the negative pairs do not have hard negatives with different utterances of the same word. This approach is used for the sole purpose of experimenting to see if each word utterance needs negative pairs from every other word utterance for the contrastive training. This also serves as a baseline for comparing the effect of hard negatives in the SSHN model.

\begin{table}[htbp]
\caption{Classification accuracy results of the models on a one vs all classification task for the words 'Four', 'Marvin' and 'Right' with the base model frozen.}
\label{tab:fdroid}
\normalsize
\centering
\begin{tabular}{ | m{0.7 cm}  |m{2 cm} |m{2 cm} | m{2 cm}|} 
\hline
\textbf{Exp} &
\textbf{Four} &
\textbf{Marvin}&
\textbf{Right}\\
\hline
C & 95.1/91.1/83.9 & 91.8/91.3/83.8 & 94.8/91.3/81.6 \\
\hline
CT1 & 98.1/97.2/93.1 & - & 93.9/92.2/88.5 \\
\hline
SC1 & \textbf{98.7/97.2/93.3} & - & 92.1/92.5/85.9 \\
\hline
CT2 &  93.2/93.1/88.9 & 82.9/77.2/76.3 & 87.9/83.7/79.5\\
\hline
SC2 & \textbf{97.9/97.1/91.6} & \textbf{93.6/92.9/83.6} & \textbf{90.9/90.7/83.1}\\

\hline
SSC & 95.1/96.3/87.9 & \textbf{88.9/89.7/81.1} & 89.7/87.7/80.9 \\
\hline
PSC & 96.7/95.7/91.5 & - & 89.6/88.4/80.2 \\
\hline
SSHN & 95.2/95.4/89.2 & - & 86.0/83.4/79.7 \\
\hline
\end{tabular}

\end{table}
\begin{table}[htbp]
\caption{Classification accuracy results of the models on a one vs all classification task for the words 'Four', 'Marvin' and 'Right'  with unfrozen base model trained with reduced learning rate.}
\label{tab:fdroid}
\normalsize
\centering
\begin{tabular}{ |m{0.7 cm} | m{2 cm} |m{2 cm}  |m{2 cm}|} 
\hline
\textbf{Exp} &
\textbf{Four} &
\textbf{Marvin}&
\textbf{Right}\\
\hline
C & 95.1/91.1/83.9 & 91.8/91.3/83.8 & 94.8/91.3/81.6 \\
\hline
CT1 & 97.6/96.4/91.1 & - & 94.8/94.1/87.8 \\
\hline
SC1 & \textbf{97.2/96.4/94.3} & - & \textbf{95.6/94.2/90.7} \\
\hline

CT2 &  97.4/96.8/90.9 & 94.2/94.7/86.5 & 95.4/92.5/87.5\\
\hline
SC2 & \textbf{98.2/98.6/93.9} & \textbf{96.3/93.2/87.3} & \textbf{95.6/94.6/89.4}\\
\hline
SSC & \textbf{97.6/98.7/89.7} & \textbf{96.0/93.9/88.6} & \textbf{96.9/94.5/87.4} \\
\hline
PSC & 98.2/97.8/92.8 & - & 94.8/94.7/88.8 \\
\hline
SSHN & 97.9/98.8/91.2 & - & 95.7/93.3/86.3 \\
\hline
\end{tabular}

\end{table}

Finally, we train a self-supervised model (SSC) with a novel training framework. This approach is aimed at solving the hard negative problems faced in the conventional self supervised training approach. Additionally it does not require a set of single word utterance dataset as in the conventional approach. In this approach, we use chunked audios from LibriSpeech long sentence audios. For this work, we use the Montreal Forced aligner tool \cite{C25} to chunk single word speech audios from the sentence audios using the available transcript. We use the transcripts only for the purpose of chunking and completely avoid it for the pre-task training. The chunking can be done using a Voice Activity Detector (VAD) in cases where the transcripts are not available. 

We use LibriSpeech [23] train-clean-100 dataset which has 100.6 hours of english sentences. We also filter the chunked audios to have a minimum size of 45KB to ensure we don't have small and repeating words like 'a', 'an' 'or' etc. We make sure that by doing the above steps most of the chunked words are unique. Thus, there is a low probability for a random pair of utterances to be the same word. 

Similar to PSC approach, for SSC we divide the chunked data randomly into set-1 and set-2  with a 30:70 ratio.  We create positive pairs from the chunked utterances and augmentations of the same utterance from set-1. The negative pairs consist of random pairs of clean/augmented chunked utterances of set-1 and set-2. We don't pair utterances within the same set as negative pairs. These conditions ensure that set-1 words have more number of both positive and negative pairings. We find that dividing the chunked data with 30:70 ratio provides better results on main task than other ratios (50:50, 70:30 etc). The bigger set-2 enables more number of distinct words for negative pairs, enabling the model to learn from more negative pairs. We also hypothesize that the set approach provides more pairs per utterance for the set-1 words aiding the model to learn the boundaries clearly.

\section{Experiments}

Table I shows the different training datasets used for the pre-task. We use batch size of 64 for all the experiments. The contrastive learning models are trained for 3 epochs on the positive and negative pairs of data. We train the C model with random initialization for 15 epochs on the classification task. For the transfer models with both the frozen and unfrozen base model, we train till the models converge to a good train/val accuracy. We find the classification transfer models converge around 8-10 epochs and the contrastive transfer models converge around 3-6 epochs for one vs all and multi-class classification.

Table II shows classification accuracy results for all the transfer models on a one vs all classification task for the words 'Four', 'Marvin' and 'Right'. The ResNet base model is frozen before the main task training in this experiment. We used clean utterances as present in the GSC dataset for the training after a 80/20 train/test split. The trained models were tested on test dataset as well as on augmentations of the test dataset. The results are represented as 'a/b/c' where a, b, c are accuracies on clean, car noise augmented and other noise (cafe, babble, music, kitchen) augmented test data. We use car noise augmentation as a specific case of uniform background noise. Table III shows the results for the same experiments with an unfrozen base model trained with reduced learning rate. The C model is same as Table II. The CT1 and SC1 models do not have results for the word 'Marvin' as it is part of the pre-task dataset for these models..

\begin{table}[htbp]
\caption{Classification accuracy results of the models on a multi trigger-word classification task for the words not used in pre-task with frozen and unfrozen base model.}
\label{tab:fdroid}

\normalsize
\centering
\begin{tabular}{  |m{0.7 cm} | m{3 cm} |m {3 cm}|} 
\hline
\textbf{Exp} &
\textbf{Frozen base} &
\textbf{Unfrozen base} \\
\hline
C     & - & 95.4/93.8/81.4  \\
\hline
CT1    & 97.6/97.1/90.6 & 97.4/97.1/91.7  \\
\hline
SC1 & 94.5/92.8/84.0 & 97.3/97.2/90.1  \\
\hline
CT2    &  95.1/95.4/86.6 & 91.0/90.6/78.4 \\

SC2 & 92.4/92.7/81.9 & \textbf{97.1/96.9/89.8} \\
\hline
SSC & 92.0/91.3/77.9 & \textbf{96.5/95.7/86.9}  \\
\hline
PSC & 93.7/93.4/83.7 & 97.5/96.9/89.0  \\
\hline
SSHN & 90.6/90.6/77.9 & 97.4/96.6/88.2  \\

\hline
\end{tabular}
\end{table}

\begin{table}[htbp]
\caption{Classification accuracy results of the models on a multi trigger-word classification task for all 35 words and a one vs all classification task for the word "backward".}
\label{tab:fdroid}
\normalsize
\centering
\begin{tabular}{ |>{\centering} m{1.6 cm} | m{2 cm} |m {2 cm}|} 
\hline
\textbf{ Exp} &
\textbf{All Data} &
\textbf{Clean Data} \\
\hline
\shortstack{C \\ (multi-class)}     & 89.3/79.07/75.2 & 84.2/80.7/60.8  \\
\hline
\shortstack{SC1 \\ (multiclass)} & 87.2/85.4/70.5 & \textbf{94.3/83.4/69.8}  \\
\hline
\shortstack{C \\ (backward)}     & 95.6/95.3/93.3 & 96.2/87.5/80.8   \\
\hline
\shortstack{SC1 \\ (backward)} & \textbf{97.4/97.8/92.9} & \textbf{96.8/96.6/92.6}   \\

\hline

\end{tabular}

\end{table}

Table IV shows accuracy of various experiments on a multi-class classification task performed with only the clean utterances of the 5 words not used in pre-task training. The results for models trained with frozen base model and unfrozen base models are shown. As in previous experiments the C model is trained with unfrozen base model with random initialization. 

Table V shows accuracy of C and SC1 models on a multi-class classification task with all 35 trigger-words (30 pre-task and 5 main task). The SC1 model is trained with a frozen base model. We perform the experiment by training with just clean data as well as with clean and augmented data. Table V also shows the accuracy of a one vs all classification task performed with the word 'backward'. This word is part of the pre-task dataset for the SC1 model. The results are recorded for a training with just clean data as well as clean and augmented data. We do these two experiments to show the generalization capacity of the contrastive models to noisy data when trained with just clean data. We don't explore further in this direction as the main motive of this paper is to test the performance on words not used in pre-task and with only clean training data.

\section{Discussions}

From results in Table II-IV,  we see that the contrastive models have comparable or improved results on the classification task. The better results from both supervised and self supervised contrastive models are in bold. The supervised contrastive models have slightly better results than the self-supervised models as seen in most other works on contrastive pre-training. Even though SC1 and CT1 have very similar performance, we see that SC2 clearly outperforms CT2 proving that the supervised contrastive learning is stable with decrease in number of trigger-words used for training. 

SSC model has comparable results to SC models in most cases. They have comparable or better results than PSC, SSHN and CT models. The SSC models are seen to perform well even though they are trained on an entirely different distribution of data compared to other models for the pre-task. The PSC vs SSHN comparison is made just to show the effect of the hard negative training. The SSHN model is out performed by the PSC model in almost all the cases.  

Table V shows that the SC model performs well in noisy conditions even when trained on only clean data. Even though the multi-class classification results are better on the classification model trained with clean and augmented data, the SC model performs better when both models are trained only with clean data . In the one vs all classification model for the word "backward" ,which is part of pre-task dataset, the SC model trained with just clean data is equivalent to a classification model trained with clean and noise augmented data. We do the above experiment to add weight to the claim that the contrastive pre-task training makes the main task models robust to noise.

The contrastive pre-task techniques (SC and SSC) can be used as a generic pre-training framework for noise robust trigger-word detection tasks in voice assistants.  We see that the SC technique is robust to noisy conditions when trained with less amounts of clean data. This reduces the need for extensive data collection and noise augmentations during the main task training thereby making it a viable option for on-device training of new trigger words. Moreover, The SC technique can be used for improving the performance of existing trigger-word detection models. Positive and negative pairs including the trigger-word can be added to the pre-task which makes the main task trigger word detection better as shown the one vs all classification case in Table V. 

The SSC technique, circumvents the need for labels in the pre-task training which makes it suitable for using the pre-task across languages without single word utterance datasets. We can use speech audios from any given language for the SSC pre-task. The learned weights can then be used for training a trigger-word detection model for words in that language with less labelled training data.

In this work we use a random split of chunked dataset for the self-supervised task. While this may ensure that there are less hard negative pairs because of the LibriSpeech dataset, it may not be completely free of hard negatives. As future work, we would like to come up with better pairing strategies for the chunked audios.

\section{Conclusions}

In this work, we show that we can get comparable results on one vs all and multi-class trigger-word classification tasks with contrastive transfer learning pre-task training. We show this in a transfer learning setting where the pre-task is trained on a different set of words from the classification main task. We also show the stability of a supervised contrastive learning task when trained with lesser trigger-word pairs. In addition, we introduce a new self-supervised training technique using random chunked words from long sentence audios without labels that counters the hard negative problem faced by conventional self-supervised techniques. We show that this technique has comparable results to other supervised and self-supervised techniques.

\bibliographystyle{IEEEtran}

\bibliography{mybib}

\end{document}